# Time operator in QFT with Virasoro constraints

Zhi-Yong Wang, Qi Qiu, Cai-Dong Xiong

E-mail:  zywang@uestc.edu.cn

*School of Optoelectronic Information, University of Electronic Science and Technology of China, Chengdu 610054, China*

**Abstract**

Time operator is studied on the basis of field quantization, where the difficulty stemming from Pauli's theorem is circumvented by borrowing ideas from the covariant quantization of the bosonic string, i.e., one can remove the negative energy states by imposing Virasoro constraints. Applying the index theorem, one can show that in a different subspace of a Fock space, there is a different self-adjoint time operator. However, the self-adjoint time operator in the maximal subspace of the Fock space can also represent the self-adjoint time operator in the other subspaces, such that it can be taken as the single, universal time operator. Furthermore, a new insight on Pauli's theorem is presented.

## 1. Introduction

Though a quantum field can be treated as a system of an infinite number of quantum particles where creation and annihilation of particles are possible, there is an uncertainty relation between the center-of-mass position and the total momentum of the system (in fact, a so-called "particle" may actually correspond to a system, such that its position and momentum are the center-of-mass variables of the system). In addition, each of field quanta as a single particle, its observables (including position and momentum) are represented by operators acting on a Hilbert space of quantum states, and its position-momentum uncertainty relation is always valid. Therefore, the fact that the spacetime coordinates of field operators play the role of parameters, does not conflict with the existence of time and position operators. For example, in many cases, time is not a mere parameter, but an intrinsic property characterizing the duration of certain physical processes (such as the lifetime of unstable particles) [1, 2].

However, according to Pauli's argument [3], in terms of a Hamiltonian $\hat{H}$ and the canonical commutation relation $[\hat{T},\hat{H}]=-i$ one define a time operator $\hat{T}$. It follows from $[\hat{H},\hat{T}^n]=in\hat{T}^{n-1}$ ( $n=1,2,3...$ ) that $[\hat{H},\exp(i\alpha\hat{T})]=-\alpha\exp(i\alpha\hat{T})$, where $\alpha$ is a real constant. Let $\hat{H}|\psi\rangle=E|\psi\rangle$, one can obtain $\hat{H}\exp(i\alpha\hat{T})|\psi\rangle=(E-\alpha)\exp(i\alpha\hat{T})|\psi\rangle$, that is, $\exp(i\alpha\hat{T})|\psi\rangle$ is also the eigenstate of the Hamiltonian $\hat{H}$ with the eigenvalue $(E-\alpha)\in(-\infty,+\infty)$. Then, the existence of the self-adjoint time operator $\hat{T}$ satisfying the canonical commutation relation $[\hat{T},\hat{H}]=-i$, contradicts the fact that the spectrum of the Hamiltonian $\hat{H}$ must be semi-bounded. Such conclusion can be called *Pauli's theorem*. As a consequence, people have presented many investigations on how to introduce time operator, how to define tunneling time, arrival time, traversal time, as well as how to describe the time–energy uncertainty relation, etc. [1, 2, 4-31].

Pauli's theorem is rigorous mathematically, it means that the existence of a self-adjoint time operator canonically conjugating to a Hamiltonian results in the appearance of negative energy states, such that the Hilbert space is bigger than the actual physical spectrum, i.e., it contains both physical and unphysical states. Nevertheless, in the covariant quantization of the Bosonic string [32, 33], the difficulty rooting in Pauli's theorem is overcome in such a way: one can sort out the physical states from the extended Hilbert space by imposing Virasoro constraints (by means of the highest weight representation of a Virasoro algebra). Moreover, the difficulty rooting in Pauli's theorem can also be overcome by means of non-self-adjoint time operators. For example, the non-self-adjoint (but hermitian, or better maximal hermitian) operator proposed in Ref. [1, 2, 4-9], as an operator for the observable time in quantum mechanics, is fully acceptable, mathematically and physically, without having to modify or split the ordinary Hilbert space.

In this paper, by borrowing ideas from the covariant quantization of the bosonic string, we will study time operator on the basis of field quantization, and will present a new insight on Pauli's theorem. We will work in natural units whereby $c = \hbar = G = 1$.

## 2. Time operator based on field quantization

For convenience, let us consider a single-mode scalar field (or a polarization component of other bosonic fields) with the frequency of $\omega$, its number operator $\hat{N} = \hat{a}^\dagger \hat{a}$ satisfies $\hat{N}|n\rangle = n|n\rangle$ ( $n = 0, 1, 2...$ ), and then its Hamiltonian $\hat{H} = (\hat{N} + 1/2)\omega$ satisfies $\hat{H}|n\rangle = E_n|n\rangle$ with $E_n = (n + 1/2)\omega$. Just as that for a different system there may correspond to a different Hamiltonian, to characterize the duration of different physical processes there may have different time operators with different physical meanings. In the present case, let us defined a time operator $\hat{T}_m$ ( $m = 1, 2...$ ) by

$$\exp(im\omega \hat{T}_m) = [\Gamma(\hat{N}+1)/\Gamma(\hat{N}+m+1)]^{1/2} \hat{a}^m, \tag{1-1}$$

$$\exp(-im\omega \hat{T}_m) = \hat{a}^{\dagger m}[\Gamma(\hat{N}+1)/\Gamma(\hat{N}+m+1)]^{1/2}, \tag{1-2}$$

where $\Gamma(x)$ is the Gama function satisfying $\Gamma(x+1) = x\Gamma(x)$ (in particular, $\Gamma(n+1) = n!$ for $n = 0, 1, 2...$ ), and the related operator function is defined according to

$$[\Gamma(\hat{N}+1)/\Gamma(\hat{N}+m+1)]^{1/2}|n\rangle = [\Gamma(n+1)/\Gamma(n+m+1)]^{1/2}|n\rangle. \tag{2}$$

The operators on the right-hand side of Eq. (1) (i.e., $[\Gamma(\hat{N}+1)/\Gamma(\hat{N}+m+1)]^{1/2} \hat{a}^m$ and its hermitian conjugate) are always meaningful, but their logarithms (offering the time operator $\hat{T}_m$) are valid only in the subspace denoted as $F_m = \{|n\rangle, n \geq m\}$ ( $m = 1, 2...$ ). In fact, using Eqs. (1) and (2), and consider that ( $p, q = 0, 1, 2...$ )

$$\hat{a}^p \hat{a}^{\dagger q}|0\rangle = \begin{cases} [q!/\sqrt{(q-p)!}]|q-p\rangle, & q \geq p \\ 0, & q < p \end{cases}, \tag{3}$$

one can prove that

$$\hat{H}\exp(im\omega\hat{T}_m)|n\rangle = \begin{cases} E_{n-m}|n-m\rangle, & m \leq n \\ 0, & m > n \end{cases}. \tag{4}$$

It follows from Eq. (4) that

$$\exp(im\omega\hat{T}_m)|n\rangle = \begin{cases} |n-m\rangle, & n \geq m \\ 0, & n < m \end{cases}. \tag{5}$$

Eq. (5) implies that the definition (1) is valid only in the subspace $F_m = \{|n\rangle, n \geq m\}$. Using $[\hat{a}, \hat{a}^\dagger] = 1$, $\hat{H} = (\hat{N} + 1/2)\omega$ and Eq. (1), one has

$$[\hat{H}, \exp(im\omega\hat{T}_m)] = -m\omega\exp(im\omega\hat{T}_m), \tag{6}$$

it follows that

$$[\hat{T}_m, \hat{H}] = -\mathrm{i}. \tag{7}$$

Likewise, the canonical commutation relation (7) is valid only in the subspace $F_m = \{|n\rangle, n \geq m\}$. Let us define operators $\hat{M}_m$'s ($m = 1, 2...$) as follows:

$$\hat{M}_m = [\hat{H}\exp(im\omega\hat{T}_m) + \exp(im\omega\hat{T}_m)\hat{H}]/2. \tag{8}$$

It is easy to prove that the operators $\hat{M}_m$'s generate the Virasoro algebra of central charge $c = 0$:

$$[\hat{M}_m, \hat{M}_n] = (m-n)\omega\hat{M}_{m+n}. \tag{9}$$

This centreless Virasoro algebra is called the Witt algebra. By means of the highest weight representation of the centreless Virasoro algebra, one can remove the dilemma set by Pauli's theorem (a more rigorous discussion will be presented in the next section). That is, in view of the fact that the Hamiltonian spectrum is on mass-shell and is bounded below, the physical states should be restricted to the set of $\{\hat{M}_m|0\rangle\}$ with the Virasoro constraints:

$$\hat{M}_0|0\rangle = E_0|0\rangle, \quad \hat{M}_n|0\rangle = 0 \ (n > 0), \tag{10}$$

where the vacuum state $|0\rangle$ plays the role of the highest weight state, $E_0 = \omega/2$ is the ground state energy of the system. Moreover, one can prove that

$$\begin{cases} \langle l|\exp(im\omega\hat{T}_m)\exp(-im\omega\hat{T}_m)|k\rangle = \delta_{kl}, \ m=1,2...;k,l=0,1,2... \\ \langle l|\exp(-im\omega\hat{T}_m)\exp(im\omega\hat{T}_m)|k\rangle = \delta_{kl}, \ m\leq k,l \text{ and } k,l,m=1,2... \end{cases}, \quad (11)$$

and then for $|k\rangle, |l\rangle \in F_m = \{|n\rangle, n \geq m\}$, one has

$$\langle l|\exp(im\omega\hat{T}_m)\exp(-im\omega\hat{T}_m)|k\rangle = \langle l|\exp(-im\omega\hat{T}_m)\exp(im\omega\hat{T}_m)|k\rangle = \delta_{kl}. \quad (12)$$

That is, in the subspace of $F_m = \{|n\rangle, n \geq m\}$, the operator $\exp(-im\omega\hat{T}_m)$ is the inverse of $\exp(im\omega\hat{T}_m)$ and *vice versa*, which implies that they are unitary operators, and it follows from Eq. (1) that $\hat{T}_m = \hat{T}_m^\dagger$, i.e., the time operator $\hat{T}_m$ given by Eq. (1) is self-adjoint in the subspace of $F_m = \{|n\rangle, n \geq m\}$. For $m > n$, $\exp(im\omega\hat{T}_m)|n\rangle = 0$ is no longer a state vector of the Hilbert space, which is equivalent to the Virasoro constraints given by Eq. (10) and ensures the absence of states with negative energy.

In fact, one can also discuss the self-adjointness of $\hat{T}_m$ on the basis of a notion of index or an index theorem. The index of an operator $\hat{\Omega}$ is defined as [34]

$$\text{index}\hat{\Omega} = \dim\ker\hat{\Omega} - \dim\ker\hat{\Omega}^\dagger. \quad (13)$$

where $\ker\hat{\Omega} = \{|\psi\rangle: \hat{\Omega}|\psi\rangle = 0\}$ is the kernel space of $\hat{\Omega}$ and $\dim\ker\hat{\Omega}$ is its dimension. One can prove that [35, 36], the operator $\hat{\Omega}$ can be decomposed as the product of an unitary operator $\hat{U}$ and a self-adjoint operator $\hat{W}$, i.e., $\hat{\Omega} = \hat{W}\hat{U}$ (or $\hat{\Omega} = \hat{U}\hat{W}$) with $\hat{U}^\dagger = \hat{U}^{-1}$ and $\hat{W}^\dagger = \hat{W}$, *if and only if* $\text{index}\hat{\Omega} = 0$. Eq. (1-1) implies that the polar decomposition

$$\hat{a}^m = [\Gamma(\hat{N}+1)/\Gamma(\hat{N}+m+1)]^{-1/2}\exp(im\omega\hat{T}_m). \quad (14)$$

Let $\hat{\Omega} = \hat{a}^m$, $\hat{U} = \exp(im\omega\hat{T}_m)$, $\hat{W} = [\Gamma(\hat{N}+1)/\Gamma(\hat{N}+m+1)]^{-1/2}$. Obviously, $\hat{W}$ is a

self-adjoint operator in the full Fock space of $\{|n\rangle, n = 0,1,2...\}$. In the subspace of $F_m = \{|n\rangle, n \geq m\}$, one has $\text{index}\hat{\Omega} = \text{index}\hat{a}^m = 0$, and then $\hat{U} = \exp(\text{i}m\omega\hat{T}_m)$ is an unitary operator, which implies that the time operator $\hat{T}_m$ is a self-adjoint operator in the subspace of $F_m = \{|n\rangle, n \geq m\}$, and in agreement with Eq. (12).

It follows from $F_m = \{|n\rangle, n \geq m\}$ that $F_1 \supseteq F_2 \supseteq F_3 \supseteq ...$. On the other hand, $\hat{T}_m$ represents the time operator of the field containing $n \geq m$ field quanta, and then $\hat{T}_1$ takes all other operators $\hat{T}_m$'s as its particular cases, which is denoted as $\hat{T}_1 \supseteq \hat{T}_2 \supseteq \hat{T}_3 \supseteq ...$. As a result, one can take the time operator $\hat{T} = \hat{T}_1$ as the single, universal time operator, it is self-adjoint in the maximal subspace of $F_1 = \{|n\rangle, n \geq 1\}$. For the moment, Eqs. (1) and (5) become as, respectively,

$$\exp(\text{i}\omega\hat{T}) = [\Gamma(\hat{N}+1)/\Gamma(\hat{N}+2)]^{1/2}\hat{a}, \ \exp(-\text{i}\omega\hat{T}) = \hat{a}^\dagger[\Gamma(\hat{N}+1)/\Gamma(\hat{N}+2)]^{1/2}. \quad (15)$$

$$\exp(\text{i}\omega\hat{T})|n\rangle = \begin{cases} |n-1\rangle, & n \geq 1 \\ 0, & n = 0 \end{cases}. \quad (16)$$

In a word, when a scalar field (or a polarization component of other bosonic fields) consists of $n \geq 1$ field quanta, its time operator $\hat{T}$ can be defined by Eq. (15), and $\hat{T}$ is self-adjoint in the subspace of $F_1 = \{|n\rangle, n \geq 1\}$. To circumvent the difficulty stemming from Pauli's theorem, one can remove all the negative-energy states by means of the quantum constraints given by Eq. (16) or Eq. (10). Moreover, the self-adjointness of $\hat{T}$ in the subspace of $F_1 = \{|n\rangle, n \geq 1\}$ can also be shown by the index theorem.

### 3. Some further considerations

For a system formed by a single-mode bosonic field with $n$ ($n \geq m = 0,1,2,...$) field quanta of frequency $\omega$, we define a general time operator by ($0 < \lambda < 1$):

$$\exp[i(m+\lambda)\omega\hat{T}_{m+\lambda}] \equiv [\Gamma(\hat{N}+1)/\Gamma(\hat{N}+m+\lambda+1)]^{1/2}\hat{a}^{\lambda}\hat{a}^m, \qquad (17\text{-}1)$$

$$\exp[-i(m+\lambda)\omega\hat{T}_{m+\lambda}] \equiv \hat{a}^{\dagger m}\hat{a}^{\dagger \lambda}[\Gamma(\hat{N}+1)/\Gamma(\hat{N}+m+\lambda+1)]^{1/2}. \qquad (17\text{-}2)$$

For $m = 0, 1, 2, ...$ and $0 \leq \lambda < 1$, $\alpha = \pm(m+\lambda)$ can represent arbitrary real number. On the other hand, for $0 < \lambda < 1$, the domains of $\exp[\pm i(m+\lambda)\omega\hat{T}_{m+\lambda}]$ do not belong to the Hilbert space formed by $\{|n\rangle, n = 0, 1, 2...\}$. However, when the system is put into a constant and uniform potential field with the potential of $V = \mp\lambda\omega$, it is equivalent to choose a new zero-energy reference point for the system via the transformation of $|n\rangle \to \exp(\pm iV\hat{T}_{m+\lambda})|n\rangle$, and then the time operator $\hat{T}_{m+\lambda}$ also plays the role of an energy-shift generator. Therefore, one can regard $\hat{T}_{m+\lambda}$ defined by Eq. (17) as the time operator of the system under the potential field of $V = -\lambda\omega$.

To study the operators $\hat{a}^{\lambda}$ and $\hat{a}^{\dagger\lambda}$ for $0 < \lambda < 1$, let us apply the Bargmann representation [37] in which one has

$$\hat{a}^{\dagger} \leftrightarrow z, \quad \hat{a} \leftrightarrow d/dz, \quad |n\rangle \leftrightarrow f_n(z) = z^n/\sqrt{\Gamma(n+1)}, \qquad (18)$$

then one has $\hat{a}^{\dagger\lambda} \leftrightarrow z^{\lambda}$. By means of fractional derivatives [38, 39], one has:

$$\hat{a}^{\lambda}|n\rangle \leftrightarrow D^{\lambda}f_n(z) = \frac{d^{\lambda}f_n(z)}{dz^{\lambda}} \equiv \frac{1}{\Gamma(1-\lambda)}\frac{d}{dz}\int_0^z \frac{f_n(t)}{(z-t)^{\lambda}}dt, \quad 0 < \lambda < 1. \qquad (19)$$

Define *virtual states* $|n\pm\lambda\rangle$ whose Bargmann representations are

$$|n\pm\lambda\rangle \leftrightarrow f_{n\pm\lambda}(z) = z^{n\pm\lambda}/\sqrt{\Gamma(n\pm\lambda+1)}, \quad 0 < \lambda < 1. \qquad (20)$$

The virtual states $|n\pm\lambda\rangle$ can describe the states of *n*-particles in a constant and uniform potential field with the potential of $V = \pm\lambda\omega$. Note that $\hat{N}^{\lambda} = (\hat{a}^{\dagger}\hat{a})^{\lambda} \neq \hat{a}^{\dagger\lambda}\hat{a}^{\lambda}$, and $\hat{a}^{\lambda}\hat{a}^m|n\rangle = \hat{a}^m\hat{a}^{\lambda}|n\rangle$ is valid only for $n \geq m$. Using

$$D^{\lambda}z^n = \frac{\Gamma(n+1)}{\Gamma(n-\lambda+1)}z^{n-\lambda}, \quad 0 < \lambda < 1, \qquad (21)$$

one can prove that, for $0 < \lambda < 1$,

$$\hat{a}^\lambda |n\rangle = \frac{\sqrt{\Gamma(n+1)}}{\sqrt{\Gamma(n-\lambda+1)}} |n-\lambda\rangle, \quad \hat{a}^{\dagger\lambda} |n\rangle = \frac{\sqrt{\Gamma(n+\lambda+1)}}{\sqrt{\Gamma(n+1)}} |n+\lambda\rangle, \qquad (22)$$

$$\hat{N} |n \pm \lambda\rangle = (n \pm \lambda) |n \pm \lambda\rangle, \qquad (23)$$

$$[\hat{H}, \exp[i(m+\lambda)\omega \hat{T}_{m+\lambda}]] = -(m+\lambda)\omega \exp[i(m+\lambda)\omega \hat{T}_{m+\lambda}], \qquad (24)$$

$$\exp[i(m+\lambda)\omega \hat{T}_{m+\lambda}] |n\rangle = \begin{cases} |n-m-\lambda\rangle, & m \leq n \\ 0, & m > n \end{cases}, \qquad (25)$$

For the moment, the Virasoro constraints become ($\alpha = m + \lambda$, $E_0 = \omega/2$):

$$\hat{M}_0 |0\rangle = E_0 |0\rangle, \quad \hat{M}_\alpha |0\rangle = 0 \text{ for } \alpha > \lambda \text{ (or } m > 0\text{)}. \qquad (26)$$

Eq. (25) or (26) preserves the semi-bounded character of the Hamiltonian spectrum. One can show that $\langle m+\lambda | n+\lambda \rangle = \langle m-\lambda | n-\lambda \rangle = \delta_{mn}$. For $\alpha = m + \lambda$ and $k, l \geq m = 0, 1, \ldots$, one has

$$\langle k | \exp(-i\alpha\omega\hat{T}_\alpha) \exp(i\alpha\omega\hat{T}_\alpha) | l \rangle = \langle k | \exp(i\alpha\omega\hat{T}_\alpha) \exp(-i\alpha\omega\hat{T}_\alpha) | l \rangle = \delta_{kl}. \qquad (27)$$

Eq. (27) shows that, as the time operator of a single-mode bosonic field with $n$ ($n \geq m$) field quanta in a potential field with the potential of $V = -\lambda \omega$, $\hat{T}_\alpha = \hat{T}_{m+\lambda}$ is self-adjoint in the subspace of $F_m = \{|n\rangle, n \geq m\}$ ($m = 0, 1, 2 \ldots$), which can also be shown by the index theorem. Because of $F_0 \supseteq F_1 \supseteq F_2 \supseteq \ldots$, one can take the time operator $\hat{T} = \hat{T}_\lambda$ ($0 < \lambda < 1$) as the single, universal time operator. For the moment, Eq. (17) becomes as

$$\exp[i\lambda\omega\hat{T}_\lambda] \equiv [\Gamma(\hat{N}+1)/\Gamma(\hat{N}+\lambda+1)]^{1/2} \hat{a}^\lambda, \qquad (28\text{-}1)$$

$$\exp[-i\lambda\omega\hat{T}_\lambda] \equiv \hat{a}^{\dagger\lambda} [\Gamma(\hat{N}+1)/\Gamma(\hat{N}+\lambda+1)]^{1/2}. \qquad (28\text{-}2)$$

It follows from Eq. (22) that $\hat{a}^\lambda |0\rangle = |-\lambda\rangle / \sqrt{\Gamma(1-\lambda)}$, and then $\hat{a}^\lambda |0\rangle \neq 0$ for $0 < \lambda < 1$, while $\lim_{\lambda \to 1} \hat{a}^\lambda |0\rangle = 0$, and $\lim_{\lambda \to 0} \hat{a}^\lambda |0\rangle = |0\rangle$. As $0 < \lambda < 1$, the virtual state $|-\lambda\rangle$ represents the vacuum state $|0\rangle$ in a potential field with the potential of $V = -\lambda\omega$.

As mentioned before, one can regard the virtual states $|n \pm \lambda\rangle$ ($0 < \lambda < 1$) as the states of *n*-particles in a potential field with the potential of $V = \pm\lambda\omega$, for which one may also present another equivalent picture: under the influence of the potential $V = \pm\lambda\omega$, the field excites or annihilates a new mode with the frequency of $\lambda\omega$, such that $|n + \lambda\rangle$ represents the state of $|n_1\rangle|1_2\rangle$ formed by $n_1$ particles of frequency $\omega_1 = \omega$ and one particle of frequency $\omega_2 = \lambda\omega$, while $|n - \lambda\rangle$ represents that a particle of frequency $\omega_2 = \lambda\omega$ is annihilated from the original field consisting of $n_1$ particles with the frequency of $\omega_1 = \omega$. In particular, if $\lambda = \lambda(t, \boldsymbol{x})$, there will be an acceleration field related to the four-dimensional (4D) gradient of $V = \pm\lambda\omega$, and one may present $|n \pm \lambda\rangle$ with a dynamical interpretation in terms of the Unruh effect (it is often stated that a uniformly accelerated observer in vacuum will 'see' thermal radiation) [40, 41], which may be developed in our next work.

### 4. A new insight on Pauli's theorem

Now, let us present a new insight on Pauli's theorem. If $|\psi\rangle$ stands for the eigenvector of a Hamiltonian operator $\hat{H}$ with the eigenvalue of $E$, then it is also the eigenvector of $\hat{H} + V$ with the eigenvalue of $E + V$, where $V$ is a potential energy. That is, one has:

$$\hat{H}|\psi\rangle = E|\psi\rangle \Rightarrow (\hat{H} + V)|\psi\rangle = (E + V)|\psi\rangle. \tag{29}$$

We call Eq. (29) *the energy-shift theorem*, it shows that the system has a degree of freedom related to its zero-energy reference point. Usually, physical observables just involve energy differences and do not depend on the choice of zero-energy reference points [24].

On the other hand, assuming all the eigenvectors $|\psi\rangle$ form a basis of a Hilbert space, for the moment $\hat{H}$ and $\hat{H} + V$ are two second-order tensors in the Hilbert space. In terms of a self-adjoint time operator $\hat{T}$ satisfying $[\hat{T},\hat{H}]=-\text{i}$, one can obtain Pauli's theorem

$$\hat{H}|\psi\rangle=E|\psi\rangle\Rightarrow\hat{H}\exp(-\text{i}V\hat{T})|\psi\rangle=(E+V)\exp(-\text{i}V\hat{T})|\psi\rangle. \tag{30}$$

To describe the energy-shift transformation of $\hat{H}\to\hat{H}+V$, there are two equivalent points of view: active transformation and passive transformation. Under the active transformation, one has $|\psi\rangle\to|\psi\rangle$, $\hat{H}\to\hat{H}'=\exp(\text{i}V\hat{T})\hat{H}\exp(-\text{i}V\hat{T})=\hat{H}+V$, and then

$$\hat{H}|\psi\rangle\to\hat{H}'|\psi\rangle=(\hat{H}+V)|\psi\rangle=(E+V)|\psi\rangle. \tag{31}$$

Under the passive transformation, one has $|\psi\rangle\to|\psi'\rangle=\exp(-\text{i}V\hat{T})|\psi\rangle$, $\hat{H}\to\hat{H}$, and then

$$\hat{H}|\psi\rangle\to\hat{H}|\psi'\rangle=\hat{H}\exp(-\text{i}V\hat{T})|\psi\rangle=(E+V)\exp(-\text{i}V\hat{T})|\psi\rangle. \tag{32}$$

The equivalence between the active and passive transformations can also be shown in terms of observable quantities, such as $\langle\psi|\hat{H}'|\psi\rangle=\langle\psi'|\hat{H}|\psi'\rangle$. In a word, Eqs. (29) and (30) present two equivalent pictures for the energy-shift theorem via the active and passive transformations, respectively. From the active perspective, the energy-shift theorem is stated as follows: if $|\psi\rangle$ stands for the eigenvector of $\hat{H}$ with the eigenvalue of $E$, then it is also the eigenvector of $\hat{H}+V$ with the eigenvalue of $E+V$; from the passive perspective, the energy-shift theorem is expressed as follows: if $|\psi\rangle$ stands for the eigenvector of $\hat{H}$ with the eigenvalue of $E$, then $\exp(-\text{i}V\hat{T})|\psi\rangle$ also stands for the eigenvector of $\hat{H}$ with the eigenvalue of $E+V$.

Therefore, the existence of the self-adjoint time operator $\hat{T}$ implies that one can present an equivalent description for Eq. (29) via Eq. (30), where Eq. (29) means that the system has a degree of freedom related to its zero-energy reference point.

## 5. Discussions and conclusions

According to Pauli's theorem, the existence of a time operator canonically conjugating to a Hamiltonian results in the appearance of negative energy states, such that the Hilbert space is bigger than the actual physical spectrum and contains both physical and unphysical states. To circumvent this difficulty, one can sort out the physical states from the extended Hilbert space by imposing Virasoro constraints (this is similar to the Gupta–Bleuler prescription in electrodynamics), in such a way one can define a self-adjoint time operator in a subspace of the Hilbert space, where the self-adjointness of the time operator in the subspace can also be shown by the index theorem. The self-adjoint time operator in the maximal subspace of a Fock space can also represent the ones in the other subspaces of the Fock space, such that it can be taken as the single, universal time operator in the Fock space. As a result, one cannot define a self-adjoint time operator for the vacuum of a quantum field, unless the quantum field is placed into a potential field.

Moreover, we have presented a new insight on Pauli's theorem, by which we reinterpret the fact that the existence of the self-adjoint time operator results in that the Hilbert space becomes bigger than the actual physical spectrum. That is, Pauli's theorem expressed by Eq. (30) is an equivalent description for the energy-shift theorem expressed by Eq. (29), and then it just implies that the physical observables of the Hamiltonian system do not depend on the choice of zero-energy reference points (Einstein equation for gravity depends on the choice of zero-energy reference points, but gravitational potential energies can be unbounded, such that in the present of gravitational fields, one can introduce a self-adjoint time operator without causing any trouble). Based on such an insight, the existence of a self-adjoint time operator is not a catastrophe, which is also in agreement with the results presented in Sections 2 and 3.

**Acknowledgements**

The first author (Z. -Y. Wang) would like to thank professors E. Recami and V. S. Olkhovsky for their helpful discussions, and thank professor J. G. Muga for his useful comments. This work was supported by the Fundamental Research Funds for the Central Universities (Grant No: ZYGX2010X013).**References**

[1] V. S. Olkhovsky, E. Recami, Lett. Nuovo Cimento (first series) 4 (1970) 1165.

[2] V. S. Olkhovsky, E. Recami, A. Gerasimchuk, Nuovo Cimento A 22 (1974) 263.

[3] W. Pauli, in: S. Flugge (Ed.), Encylopaedia of physics, Vol. 5, Springer, Berlin, 1958, p. 60.

[4] V. S. Olkhovsky, E. Recami, J. Jakiel, Phys. Rep. 398 (2004) 133.

[5] V. S. Olkhovsky, E. Recami, Int. J. Mod. Phys. B 22 (2008) 1877.

[6] E. Recami, V. S. Olkhovsky, S. P. Maydanyuk, Int. J. Mod. Phys. A 25 (2010) 1785.

[7] E. Recami, A time operator and the time-energy uncertainty relation, in The Uncertainty Principle and Foundation of Quantum Mechanics, eds. C. Price and S. Chissik, J. Wiley, London, 1977, pp. 21–28.

[8] V. S. Olkhovsky, Physics-Uspekhi 54 (2011) 829.

[9] V. S. Olkhovsky, Time as a quantum observable, LAP-LAMBERT Academic Publishing, Germany, 2012.

[10] H. A. Fertig, Phys. Rev. Lett. 65 (1990) 2321.

[11] R. Landauer, Th. Martin, Rev. Mod. Phys. 66 (1994) 217.

[12] A. M. Steinberg, Phys. Rev. Lett. 74 (1995) 2405.

[13] R. Giannitrapani, Int. J. Theor. Phys.36 (1997) 1575.

[14] J. G. Muga, C. R. Leavens, J. P. Palao, Phys. Rev. A 58 (1998) 4336.

[15] I. L. Egusquiza, J. G. Muga, Phys. Rev. A 61 (1999) 012104.

[16] M. Skulimowski, Phys. Lett. A 297 (2002) 129-136; Phys. Lett. A 301 (1999) 361.

[17] E. A.Galapon , R. F. Caballar, R. T. Bahague, Phys. Rev. Lett. 93 (2004) 180406.

[18] G. Gour, F. C. Khanna, M. Revzen, Phys. Rev. A 69 (2004) 014101.

[19] P. Bokes, Phys. Rev. A 83 (2011) 032104.

[20] N. Yamada, Phys. Rev. Lett. 83 (1999) 3350.

[21] J. G. Muga, A. Ruschhaupt, A. del Campo (Eds.), Time in Quantum Mechanics,vol. 2,
12